\documentclass{JHEP3}
\usepackage{amsmath,times,amssymb,graphicx,slashed,undertilde,epsfig}
\newenvironment{narrow}[2]{
   \begin{list}{}{
   \setlength{\topsep}{0pt}
   \setlength{\leftmargin}{#1}
   \setlength{\rightmargin}{#2}
   \setlength{\listparindent}{\parindent}
   \setlength{\itemindent}{\parindent}
   \setlength{\parsep}{\parskip}}
   \item[] }{\end{list}}
\title{Scattering on the Moduli Space of $\mathcal{N}=4$ \\ Super Yang-Mills}
\author{Robert M. Schabinger \footnote{schabr@u.washington.edu}\\ Department of Physics, University of Washington, Seattle, WA ~98195-1560}

\abstract{We calculate one-loop scattering amplitudes in $\mathcal{N}=4$ super Yang-Mills theory away from the origin of the moduli space and demonstrate that the results are extremely simple, in much the same way as in the conformally invariant theory. Specifically, we consider the model where an $SU(2)$ gauge group is spontaneously broken down to $U(1)$. The complete component Lagrange density of the model is given in a form useful for perturbative calculations. We argue that the scattering amplitudes with massive external states deserve further study. Finally, our work shows that loop corrections can be readily computed in a mass-regulated $\mathcal{N}=4$ theory, which may be relevant in trying to connect weak-coupling results with those at strong coupling, as discussed recently by Alday and Maldacena.}
\begin{document}

\section{Introduction}
The Lagrange density of $\mathcal{N}=4$ super Yang-Mills theory was first written down long ago \cite{BSS} and, shortly thereafter, the first one-loop scattering amplitude in the model was calculated \cite{GSB}. Since then, such scattering amplitudes have been extensively studied by many groups (see \textit{e.g.} \cite{BDDK,BCF}). $\mathcal{N}=4$ SYM, however, has non-trivial dynamics \cite{F}. The theory possesses a moduli space of vacua parametrized by the vacuum expectation values (VEVs) of the three scalar and three pseudo-scalar fields in the model. It should be stressed that, in going to a generic point in the moduli space, the $\mathcal{N}=4$ supersymmetry will be preserved, but the gauge group of the theory will be spontaneously broken. The theory described in \cite{BSS} is the conformal phase of $\mathcal{N}=4$ SYM, where all the VEVs are equal to zero. So far, scattering amplitudes away from the origin of the moduli space (for states in the Coulomb phase of the theory) have received relatively little attention.

 $\mathcal{N}=4$ SYM is a very special four dimensional quantum field theory and, consequently, its S-matrix has several unusual properties. We begin by reviewing the interesting features of weak-coupling perturbation theory in the conformal theory. The field content of the model consists of a gauge field $A_{\mu}$, four Majorana fermions $\psi_i$, three real scalars $X_p$, and three real pseudo-scalars $Y_q$. All fields are in the adjoint representation of a compact gauge group, $G$. In this work, we choose $G = SU(2)$ for simplicity. In this case, a generic field, $\phi$, may be written in terms of its color components as $\phi = \frac{\phi_a}{2}~\sigma_a$, where $\sigma_a$ are the usual Pauli matrices. The Lagrange  density of $\mathcal{N}=4$ is given by \cite{CV} \footnote{The Lagrange density given in \cite{CV} is a superset of that for $\mathcal{N}=4$ SYM reproduced here. \cite{CV} follows \cite{W}, but corrects several misprints which exist in that reference.}
\begin{align}
\mathcal{L} &= -\textrm{tr} \bigg\{ \frac{1}{2} F_{\mu \nu} F^{\mu \nu} + \bar{\psi_i} \slashed{D}  \psi_i + D^{\mu} X_p D_{\mu} X_p + D^{\mu} Y_q D_{\mu} Y_q \\ \nonumber
& + i g \bar{\psi_i} \alpha^p_{i j}[X_p,\psi_j]-g \bar{\psi}_i \gamma_5 \beta^q_{i j} [Y_q,\psi_j] \\ \nonumber
& -\frac{g^2}{2}\bigg([X_l,X_k][X_l,X_k]+[Y_l,Y_k][Y_l,Y_k]+2[X_l,Y_k][X_l,Y_k]\bigg )\bigg\},
\end{align}
where the $4 \times 4$ matrices $\alpha^p$ and $\beta^q$ are given by  \footnote{$\sigma_0$ is the $2 \times 2$ identity matrix.}
\begin{align}
& \alpha^1 = \left(\begin{array}{cc} i \sigma_2 & 0 \\ 0 & i \sigma_2 \end{array}\right),~~\alpha^2 = \left(\begin{array}{cc} 0 & - \sigma_1 \\ \sigma_1 & 0 \end{array}\right),~~\alpha^3 = \left(\begin{array}{cc} 0 & \sigma_3 \\ -\sigma_3 & 0 \end{array}\right),  \\ \nonumber
& \beta^1 = \left(\begin{array}{cc} -i \sigma_2 & 0 \\ 0 & i \sigma_2 \end{array}\right),~~\beta^2 = \left(\begin{array}{cc} 0 & -i \sigma_2 \\ -i \sigma_2 & 0 \end{array}\right),~~\beta^3 = \left(\begin{array}{cc} 0 & \sigma_0 \\ -\sigma_0 & 0 \end{array}\right).
\end{align}
Once the gauge group and coupling constant $g$ are fixed, the theory is uniquely specified. 

It is by now well known that $\mathcal{N}=4$ SYM scattering amplitudes are ultraviolet-finite \cite{SW,HSW,Ma,BLN,HST}. A modern characterization of UV finiteness in gauge theories with references to the older literature is found in \cite{LSSVV}. In the conformal phase, all the external legs in a given scattering process are massless. In general, one-loop scattering amplitudes in massless quantum field theories may be written in terms of a basis consisting of certain bubble, triangle, and box scalar Feynman integrals with clusters of subsets of the $n$ external momenta exiting each vertex\cite{PV,BDKrev,vNV,Mel}. In $\mathcal{N}=4$ SYM a direct calculation of the effective action shows that the $n$-gluon one-loop scattering amplitude can be written in terms of scalar box Feynman integrals only. \cite{BDDK,GGRS}

This result is somewhat surprising, even in view of the fact that in $\mathcal{N}=4$ SYM scattering amplitudes all UV divergences must cancel. Scalar box and triangle integrals with no internal masses have divergences, but only in the infrared, whereas the scalar bubble integral with no internal masses has only an ultraviolet divergence. Thus, we would naively expect the one-loop $n$-gluon amplitude to contain both boxes and triangles. It turns out that this naive expectation is wrong and all the scalar triangle integrals cancel out as well. By deriving the $\mathcal{N}=4$ analog (see  \cite{BDDPR}) of the well known $\mathcal{N}=1$ supersymmetric Ward identities \cite{GP, GPvN, D} and applying them to the $n$-gluon scattering amplitudes, it can be shown that a much larger class of $\mathcal{N}=4$ amplitudes must be pure scalar box integrals as well. In particular, all four-point and five-point amplitudes fall into this category.

A natural question is how much of this interesting structure is preserved when we go to the Coulomb phase of the theory. There the situation is somewhat different because, in the Coulomb phase, we must distinguish scattering amplitudes which have only massless external states from those that include some states from massive sector of the theory. Before dicussing this, let us specify a convenient point in the moduli space to study. We give the scalar $X_1$ a vacuum expectation value of $\frac{v}{2}~ \sigma_3$:
\begin{align}
\langle X_1 \rangle = \frac{v}{2} ~\sigma_3.
\end{align}

Giving the $X_1$ a VEV in the prescribed manner has several consequences. The gauge group $SU(2)$ is spontaneously broken to $U(1)$ and the $SU(2)$ gauge multiplet of $\mathcal{N}=4$ is broken up into a $U(1)$ gauge multiplet of $\mathcal{N}=4$ and a massive vector multiplet of $\mathcal{N}=4$ charged under the $U(1)$. In the case of a spontaneously broken $SU(2)$ symmetry, there is nothing special about the choice we made for the VEV of $X_1$. This is because the symmetry breaking pattern $SU(2) \rightarrow U(1)$ is unique. For more complicated gauge groups, it is possible to obtain more than one symmetry breaking pattern. 

The $U(1)$ gauge multiplet is composed of the unbroken $U(1)$ gauge field, $A_{\mu}^0$, the four Goldstone spinors of superconformal symmetry (all four superconformal symmetries are spontaneously broken), $\psi_i^0$, the five Goldstone bosons of R-symmetry (the R-symmetry is spontaneously broken from $SO(6)$ with fifteen generators to $SO(5)$ with ten generators \cite{F}), $\{X_2^0,X_3^0,Y_1^0,Y_2^0,Y_3^0\}$, and the Goldstone boson of dilatations, $X_1^0$. The massive vector multiplet is composed of a complex vector boson, $A_{\mu}^{\pm}$, four Dirac spinors, $\psi_i^{\pm}$, and five complex scalars, $\{X_2^{\pm},X_3^{\pm},Y_1^{\pm},Y_2^{\pm},Y_3^{\pm}\}$. Each of these complex fields acquires a mass squared $m^2 = g^2 v^2$ after spontaneous symmetry breaking.

Though it is not obvious, there is evidence that each of the noteworthy properties that the $\mathcal{N}=4$ S-matrix possessed in the conformal phase carries over to the Coulomb phase. That the scattering amplitudes are still free of UV divergences follows from the arguments in \cite{LSSVV}. Whether the scattering amplitudes are all expected to be pure scalar box is still somewhat speculative. It was, however, shown fairly recently \cite{BIP} that the complete low energy effective action of $SU(2)$ $\mathcal{N}=4$ SYM broken to $U(1)$ is consistent with the amplitudes being pure box. This was done via an $\mathcal{N}=2$ superspace calculation, where \cite{BIP} allowed for background hyper-multiplets as well as background gauge multiplets. It should be stressed that the results of \cite{BIP} do not constitute a proof that the Coulomb phase amplitudes are pure box, since the effective action obtained in \cite{BIP} is only valid if all Mandelstam invariants are small relative to $m$.

 In fact, to the author's knowledge, the work of \cite{BIP} is the first paper which attempts to treat the complete $SU(2)\rightarrow U(1)$ $\mathcal{N}=4$ model; the other known treatments \cite{DS,KM} focus on the massless sector of the spontaneously broken theory. Of these works, only \cite{KM} attempts to compute a scattering amplitude. They calculate the four $X_2^0$ one-loop amplitude in the $\mathcal{N}=1$ supergraph formalism. The superspace approach is not well-suited for calculation of four-point functions, due to the fact that is not straightforward to see the underlying simplicity of results obtained in this way (see e.g. \cite{K}). 
It should be stressed, however, that the spirit of their calculation was quite prescient. The primary motivation of \cite{KM} was to study $\mathcal{N}=4$ supersymmetric scattering amplitudes with the IR divergences regulated in a natural way. Introducing a mass-regulator via spontaneous breaking of gauge symmetry is actually one of the methods by which Alday and Maldacena \cite{AM} regularized the IR divergences of the four-point gluon scattering amplitude in $\mathcal{N}=4$ at \textit{strong} coupling. 
 
 Another interesting paper \cite{ST} rederives, in a slightly different context, the well-known result \cite{GP,GPvN} that the massless sector of our model should have exactly the same supersymmetric Ward identites between scattering amplitudes as conformal $\mathcal{N}=4$. The authors of \cite{ST} point out that useful supersymmetric Ward identities must still exist when some of the external scattering states are massive. In fact, such Ward identities have already been applied to relate amplitudes with massive quarks to known amplitudes with massive scalars \cite{SWein}.
 
In this work we calculate a number of four-point one-loop scattering amplitudes in the \\ $SU(2)\rightarrow U(1)$ $\mathcal{N}=4$ model. We examine both the case where all the external legs are massless and the case where some of them are massive. In particular, we provide evidence that the scattering amplitudes with massive external states are, in fact, pure box. We focus on how the symmetries of our model constrain the answers obtained and, in some cases, allow us to relate distinct scattering amplitudes to each other. Finally we explain how our four photon one-loop scattering amplitude in the $SU(2)\rightarrow U(1)$ $\mathcal{N}=4$ model can be thought of as a mass-regulated, color-ordered, amplitude in an unbroken $SU(2)$~$ \mathcal{N}=4$ SYM theory. In other words, our work provides a weak-coupling analog of the mass regulator introduced by Alday and Maldacena \cite{AM} to compute gluon scattering amplitudes at strong coupling. 

The plan of this paper is as follows. In Section Two, we quantize the Lagrange density \\(Eqn. 1.1) in  $R_{\xi}$ gauge, motivate an efficient choice of gauge for the computations we want to do, and address a subtlety in the Feynman rules. In Section Three, we present and discuss the calculation of several one-loop four-point scattering amplitudes in both the massless and massive sectors of $\mathcal{N}=4$ SYM in the Coulomb phase. In Section Four, we summarize our results and present some ideas for future work. In Appendix A, we provide the definitions of the master integrals we use in Section Three. In Appendix B, we provide the complete Lagrange density of our model expanded in a form where the Feynman rules can  easily be derived. Finally, in Appendix C, we give the diagrammatic expansions of all one-loop amplitudes calculated in Section Three.
\section{Setup}

~~~~~~~We employ the metric diag$(-,+,+,+)$ and introduce the notation
\begin{align}
\phi^{\pm} \equiv \frac{\phi_1 \pm i \phi_2}{\sqrt{2}}
\end{align}
where $\phi$ is a generic field. In order to do perturbative calculations in our model, we must perform $R_\xi$ quantization on the classical Lagrange density given in the previous section with the VEV of Eqn. 1.3 . This is a standard calculation and the details will not be shown here. We now give those terms which were either affected by or introduced by the Fadeev-Popov procedure. These terms include those that lead to the propagators for the gauge field, $A_{\mu}^0$, the massive vector fields, $A_{\mu}^{\pm}$, the Goldstone fields, $X_1^{\pm}$, the massless Higgs-like field $X_1^0$, and the Fadeev-Popov ghosts $\{c^0,c^{\pm}\}$. Also included are the interactions between the ghosts, Higgs, Goldstones, and gauge fields:
\begin{align}
\mathcal{L}_{FP;2}=& -\frac{1}{2} A_{\mu}^0 \Big(-g^{\mu \nu} \partial^2+\partial^{\mu}\partial^{\nu} \Big( 1-\frac{1}{\xi}\Big) \Big)A_{\nu}^0-A_{\mu}^+ \Big( -g^{\mu \nu} \partial^2+\partial^{\mu}\partial^{\nu} \Big(1-\frac{1}{\xi}\Big)+m^2 g^{\mu \nu}\Big)A_{\nu}^- \nonumber \\& -X_1^+\Big(-\partial^2+ \xi m^2\Big)X_1^- -\frac{1}{2} X_1^0\Big(-\partial^2\Big)X_1^0  -\bar{c}\Big(-\partial^2+ \xi m^2\Big)c-\frac{1}{2} \bar{c}^0\Big(-\partial^2\Big)c^0 \\
\mathcal{L}_{FP;3}=& -\xi m g X_1^0 \bar{c} c+\xi m g X_1^- \bar{c} c^0 -i g \bar{c} \partial^{\mu} \big(A^0_{\mu} c \big)+i g \bar{c} \partial^{\mu} \big(A_{\mu}^- c^0 \big)+i g \bar{c}^0 \partial^{\mu} \big(A^+_{\mu} c \big) + \textrm{h.c.}
\end{align}
The interactions with the ghost fields are not relevant to the calculations performed in this paper, but are included for completeness. The propagators, on the other hand, are obviously quite important.
\begin{align}
\langle A_\mu^0(l)A_\nu^0(l) \rangle = \frac{- i \Big( g^{\mu \nu} - \frac{l^{\mu} l^{\nu} (1-\xi)}{l^2}\Big)}{l^2} && \langle A_\mu^+(l)A_\nu^-(l) \rangle = \frac{- i \Big( g^{\mu \nu} - \frac{l^{\mu} l^{\nu} (1-\xi)}{l^2+\xi m^2}\Big)}{l^2+m^2} \nonumber \\
\langle X_1^+(l)X_1^-(l) \rangle = \frac{- i}{l^2+\xi m^2} && \langle X_1^0(l)X_1^0(l) \rangle = \frac{- i}{l^2} \nonumber \\ \langle \bar{c}(l)c(l) \rangle = \frac{- i}{l^2+\xi m^2} && \langle \bar{c}^0(l)c^0(l) \rangle = \frac{- i}{l^2}
\end{align}

Traditionally, one-loop gauge theory computations have been performed in 't Hooft-Feynman gauge. It has been known, however, at least since the work of \cite{BDDK}, that working in unitary gauge ($\xi \rightarrow \infty$) has significant practical advantages in theories without UV divergences, like $\mathcal{N}=4$. The reason for this is easily understood; in $\mathcal{N}=4$ on the Coulomb branch, one only has to compute the box diagrams that arise and keep the pieces of them which are pure scalar box, since everything else must cancel in the end\footnote{This has not conclusively been shown for all scattering amplitudes with more than six external legs in the conformal phase and is purely speculative for amplitudes in the Coulomb phase which include massive external states.}. If, for example, one wanted to compute the four Higgs (denoted $X_1^0 X_1^0 X_1^0 X_1^0$ hereafter) one-loop amplitude in 't Hooft-Feynman gauge, there would be eight independent box diagrams to evaluate, whereas in unitary gauge, there would only be three\footnote{Technically, the ghost loop would be non-zero in this case also, since the coupling of $X_1^0$ to the charged ghosts carries an explicit factor of $\xi$. This, however, would not contribute a scalar box integral but only a momentum independent constant, which would play a role in the cancellation of the UV divergences if we explicitly kept track of them.}.

In other words, the fact that we have to perform relatively little integral reduction in the computation of one-loop four-point functions in $\mathcal{N}=4$, means that having fewer diagrams to evaluate benefits us more than having Feynman rules which are renormalizable by power-counting. The rest of the Lagrangian can be obtained in a straightforward (but tedious) way by making the shift $X_1 \rightarrow X_1 + \langle X_1 \rangle$ in Eqn. 1.1 and performing the traces over the $SU(2)$ matrices. The many interaction terms obtained in this way are presented in Appendix B in a form where the Feynman rules can be read off. 

There is one somewhat counter-intuitive result that arises in carrying out this program. Let us consider the derivation of the propagators of the four Dirac fermions. Upon performing the shift $X_1 \rightarrow X_1 + \langle X_1 \rangle$, we arrive at a fermion mass term $-i g ~\textrm{tr}\{ \bar{\psi}_i \alpha^1_{ij} [\langle X_1 \rangle,\psi_j]\}$ from the Yukawa interaction of the $SU(2)$ Majorana fields with $X_1$. Expanding this out and adding in the Dirac fields' kinetic terms gives
\begin{align}
\mathcal{L}_{\bar{\psi}\psi} = -\bar{\Phi}_i \slashed{\partial} \Phi_i-i m ( \bar{\Phi}_1 \Phi_2 - \bar{\Phi}_2 \Phi_1 + \bar{\Phi}_3 \Phi_4  - \bar{\Phi}_4 \Phi_3),
\end{align}
where we have introduced the notation 
\begin{align}
\bar{\psi_i}^+ \equiv \bar{\Phi}_i & & \psi_i^- \equiv \Phi_i.
\end{align}
We can now diagonalize the mass matrix in flavor space via the unitary transformation
\begin{align}
 \left(\begin{array}{c} \Phi_1 \\ \Phi_2 \end{array}\right) = \frac{1}{\sqrt{2}} \left(\begin{array}{cc} i  & - i \\ 1 & 1  \end{array}\right) \left(\begin{array}{c} \utilde{\Phi}_1 \\ \utilde{\Phi}_2 \end{array}\right).
\end{align}
After making this change of variables, the quadratic Lagrange density for the Dirac fields reads
\begin{align}
\mathcal{L}_{\bar{\psi}\psi}^{'} = -\utilde{\bar{\Phi}}_i \slashed{\partial} \utilde{\Phi}_i-m \big( \utilde{\bar{\Phi}}_1 \utilde{\Phi}_1 - \utilde{\bar{\Phi}}_2 \utilde{\Phi}_2 + \utilde{\bar{\Phi}}_3 \utilde{\Phi}_3  - \utilde{\bar{\Phi}}_4 \utilde{\Phi}_4\big),
\end{align}
which results in non-standard propagators for the $\utilde{\Phi}_2$ and $\utilde{\Phi}_4$ fields:
\begin{align}
\langle \utilde{\bar{\Phi}}_1(l) \utilde{\Phi}_1(l) \rangle = \frac{- i \big( -i \slashed{l}+m \big)}{l^2+m^2} && \langle \utilde{\bar{\Phi}}_2(l) \utilde{\Phi}_2(l) \rangle = \frac{- i \big( -i \slashed{l}-m \big)}{l^2+m^2} \nonumber \\
\langle \utilde{\bar{\Phi}}_3(l) \utilde{\Phi}_3(l) \rangle = \frac{- i \big( -i \slashed{l}+m \big)}{l^2+m^2} && \langle \utilde{\bar{\Phi}}_4(l) \utilde{\Phi}_4(l) \rangle = \frac{- i \big( -i \slashed{l}-m \big)}{l^2+m^2}.
\end{align}
If we erroneously tried to give standard Dirac propagators to all four fields, we would be able to find, amongst other things, one-loop four-scalar scattering amplitudes which violate supersymmetry. This $\utilde{\Phi}$ basis is a convenient one to use for perturbative calculations. 
\section{One-Loop Four-Point Scattering Amplitudes}
~~~~~~Before attempting to calculate an amplitude with massive external states, we first rederive a known result in the massless sector\cite{GPR} and two amplitudes related to it by known supersymmetric Ward identites\cite{BDDPR,ST}. The work of \cite{GPR} gives all independent one-loop helicity amplitudes for the (SM or minimal SUSY) process $0  \rightarrow  \gamma \gamma \gamma \gamma$. (It is convenient to consider all particles as outgoing, and we adopt this convention throughout.) This requires them to calculate the effect of a massive scalar loop, a massive fermion loop, and a massive vector loop. This is actually much more than we need to do for the calculation in $\mathcal{N}=4$. The diagrammatic representations of the amplitudes discussed in this section are given in Appendix C.

First of all, as discussed above, we may neglect all terms in the results of \cite{GPR} which do not correspond to box integrals. Furthermore, two well-known supersymmetric Ward identities are $\mathcal{M}(k_1^+,k_2^+,k_3^+,k_4^+)=0$ and $\mathcal{M}(k_1^+,k_2^+,k_3^+,k_4^-)=0$. It is easily seen by examining the proofs of these relations \cite{D} or the explicit calculations in \cite{BM} that these relations hold independent of whether the particles running in the loops are massive. These SUSY relations immediately tell us that two of the independent one-loop $A_{\mu}^0 A_{\nu}^0 A_{\rho}^0 A_{\sigma}^0$ amplitudes are identically zero to all orders in perturbation theory. It follows that $\mathcal{M}(k_1^+,k_2^+,k_3^-,k_4^-)$ is the only helicity amplitude which must be calculated. In adding up the box coefficients given in \cite{GPR}, it is important to remember that there are five scalar loops, four fermion loops, and one vector loop\footnote{See Appendix C for details.}. Adding up the pieces gives the final result 
\begin{align}
\mathcal{M}(k_1^+,k_2^+,k_3^-,k_4^-) = 8 g^4 s^2 \Big( I^{(4)}_0 (s,t)+I^{(4)}_0 (s,u)+I^{(4)}_0 (t,u) \Big),
\end{align}
where we adopt the conventions
\begin{align}
s = (k_1+k_2)^2 & & t = (k_1+k_4)^2 & & u = (k_1+k_3)^2
\end{align}
and the scalar box integrals $I^{(4)}_0 (s,t)$, $I^{(4)}_0 (s,u)$, and $I^{(4)}_0 (t,u)$ are defined in Appendix A. To calculate the helicity amplitude we used the standard non-covariant basis for the polarization vectors, where one works in the center-of-mass frame and expresses all non-zero dot products of the polarization vectors with each other and the external momenta in terms of $s$,~$t$, and $u$.

To reproduce the result of Eqn. 3.1, we worked in unitary gauge, as explained in Section Two, and employed the Mathematica package FeynCalc \footnote{FeynCalc is a framework for performing perturbative calculations in gauge theories.} \cite{M}.  An interesting feature of the calculation is that the loop momentum polynomials explicitly cancel amongst the various components. In other words, for this particular calculation, the integral reduction of the box graphs is trivial if one adds up the components first.

To illustrate this point, we give the loop momentum numerators of the (properly weighted) components which lead to the coefficient of $I^{(4)}_0 (s,t)$:
\begin{align}
 5 \times \textrm{N}\bigg( \mathcal{M}^{\textrm{scalar}}(k_1^+,k_2^+,k_3^-,k_4^-)\bigg) &= 160~ l \cdot \epsilon^+(k_1) l \cdot \epsilon^+(k_2) l \cdot \epsilon^-(k_3) l \cdot \epsilon^-(k_4) \\
 4 \times \textrm{N}\bigg( \mathcal{M}^{\textrm{fermion}}(k_1^+,k_2^+,k_3^-,k_4^-)\bigg) &= 32~s~l\cdot \epsilon^+(k_1) l\cdot \epsilon^+(k_2) + 32~s~l \cdot \epsilon^-(k_3) l \cdot \epsilon^-(k_4) \nonumber \\ &-256~ l \cdot \epsilon^+(k_1) l \cdot \epsilon^+(k_2) l \cdot \epsilon^-(k_3) l \cdot \epsilon^-(k_4) \\
\textrm{N}\bigg( \mathcal{M}^{\textrm{vector}}(k_1^+,k_2^+,k_3^-,k_4^-)\bigg) &= 8 s^2 + 96~ l \cdot \epsilon^+(k_1) l \cdot \epsilon^+(k_2) l \cdot \epsilon^-(k_3) l \cdot \epsilon^-(k_4) \nonumber \\
&- 32~s~l\cdot \epsilon^+(k_1) l\cdot \epsilon^+(k_2) - 32~s~l \cdot \epsilon^-(k_3) l \cdot \epsilon^-(k_4).
\end{align}
Adding up Eqns. 3.3-3.5 gives $8 s^2$, which immediately lets us read off the $I^{(4)}_0 (s,t)$ contribution to Eqn. 13 above.  This additional cancellation structure between the numerator loop momentum polynomials of diagrams with different internal structure is related to the fact that the external states are gauge fields. This can be understood by thinking about the calculation in  background field gauge \cite{BDDK}. 

In fact, the same cancellation structure observed above is present also in the case of the  $A_{\mu}^0 A_{\nu}^0 X_2^0 X_2^0$  amplitude\footnote{It should be stressed that we choose $X_2^0$ for concreteness. As will be discussed below, the $\mathcal{N}=4$ supersymmetry of the model demands that, alternatively, we could have replaced $X_2^0$ with any of $\{X_1^0,X_3^0,Y_1^0,Y_2^0,Y_3^0\}$ and arrived at the result of Eqn. 3.6.}. Note that, in this case, there is a supersymmetric Ward identity which sets the amplitudes with identical helicities to zero. For the non-zero helicity configurations, we find
\begin{align}
\mathcal{M}(k_1^+,k_2^-) = \mathcal{M}(k_1^-,k_2^+) = 8 g^4 t u \Big( I^{(4)}_0 (s,t)+I^{(4)}_0 (s,u)+I^{(4)}_0 (t,u) \Big).
\end{align}

An interesting point is the following. Suppose we took the result for the $A_{\mu}^0 A_{\nu}^0 A_{\rho}^0 A_{\sigma}^0$ amplitude and tried to use supersymmetric Ward identities to predict the answer for the $A_{\mu}^0 A_{\nu}^0 X_2^0 X_2^0$ amplitude. The prediction of the supersymmetric Ward identities in the conformal phase must be identical to that considered here, since the identities are independent of color and the difference between the two calculations shows up only in the basis of integrals, not in the coefficients. In other words, it should be possible to show using only $N=4$ supersymmetry that 
\begin{align}
\mathcal{M}(k_1^+,k_2^+,k_3^-,k_4^-)= \frac{s^2}{t u} \mathcal{M}(k_1^+,k_2^-)
\end{align}

It is clear from the $\mathcal{N}=4$ supersymmetry that there must be some proportionality between the two amplitudes. To find the proportionality constant, it is enough to consider the usual leading-color tree amplitudes in unbroken $\mathcal{N}=4$, due to the color-independent and non-perturbative nature of the supersymmetric Ward identites. The leading-color tree amplitude for four gluon scattering is
\begin{align}
\mathcal{A}(k_1^+,k_2^+,k_3^-,k_4^-)= \frac{\langle 3,4 \rangle^4}{\langle 1,2 \rangle \langle 2,3 \rangle \langle 3,4 \rangle \langle 4,1 \rangle}
\end{align}
and the leading-color tree amplitude for two gluon two scalar scattering is
\begin{align}
\mathcal{A}(k_1^+,k_2^-)= \frac{\langle 2,3 \rangle^2 \langle 2,4 \rangle^2}{\langle 1,2 \rangle \langle 2,3 \rangle \langle 3,4 \rangle \langle 4,1 \rangle}.
\end{align}
The ratio of these two trees is
\begin{align}
\frac{\mathcal{A}(k_1^+,k_2^+,k_3^-,k_4^-)}{\mathcal{A}(k_1^+,k_2^-)} = \frac{\langle 3,4 \rangle^4}{\langle 2,3 \rangle^2 \langle 2,4 \rangle^2}= \textrm{C} ~ \frac{s^2}{t u},
\end{align}
where C is an unimportant overall phase. 

This analysis illustrates that any other four-point amplitude in the massless sector of our model must have mass-independent coefficients, since exactly the same structures would appear if we were in the conformal phase of the theory and we have already seen that the loop integral coefficients of $\mathcal{M}(k_1^+,k_2^+,k_3^-,k_4^-)$ and $\mathcal{M}(k_1^+,k_2^-)$ have no dependence on $m$. This is a direct consequence of the fact that the supersymmetric Ward identities are the same in both cases, independent of whether the particles running in the loops acquire mass. 

Before leaving the massless sector, we calculate the one-loop $X_2^0 X_2^0 X_2^0 X_2^0$ amplitude and find a result that disagrees with that of \cite{KM}. As before, an analysis based on the supersymmetric Ward identities predicts a unique answer for the four scalar amplitude as a rational function of $s$, $t$, and $u$ times the four photon amplitude. In other words, supersymmetry tells us we could equally well choose to compute the amplitude with four of any of the other massless scalars, $\{X_1^0,X_3^0,Y_1^0,Y_2^0,Y_3^0\}$, since, in the supersymmetric Ward identity argument, the flavor of the scalar being scattered was irrelevant. 

We now address a point which was glossed over in the previous calculation. Since \\ $\{X_2^0,X_3^0,Y_1^0,Y_2^0,Y_3^0\}$ transform as a $\utilde{\textbf{5}}$ of the manifest $SO(5)$ R-symmetry, the $X_2^0 X_2^0 X_2^0 X_2^0$, $X_3^0 X_3^0 X_3^0 X_3^0$, or any of the $Y_q^0 Y_q^0 Y_q^0 Y_q^0$ amplitudes look exactly the same gauge-invariant component by gauge-invariant component\footnote{It is, however, instructive to compare the fermionic loop graphs with external scalars to those with pseudo-scalars using the Lagrange density given in Appendix B.}. What we mean is that the sum of the graphs with fermionic loops will be the same for each field in the R-symmetry multiplet and the sum of the graphs with bosonic loops will be the same for each field in the R-symmetry multiplet. 

On the other hand, it is not entirely clear from the outset how the $X_1^0 X_1^0 X_1^0 X_1^0$ amplitude works out to be identical with the other massless scalar four-point amplitudes. The difficulty is that $X_1^0$ is a singlet of $SO(5)$ and its interactions are rather different than those of the other massless scalars. To clarify this, we give the gauge-invariant components for both situations and show that they lead to the same result. We find 
\begin{align}
\mathcal{M}_{\utilde{\textbf{5}}}^{\textrm{Bose loops}} &= 8 g^4 (s^2+t^2) I^{(4)}_0 (s,t)+ 8 g^4 (s^2+u^2)I^{(4)}_0 (s,u) \nonumber \\ & + 8 g^4 (t^2+u^2)I^{(4)}_0 (t,u) \\
\mathcal{M}_{\utilde{\textbf{5}}}^{\textrm{Fermi loops}} &= 8 g^4 s t I^{(4)}_0 (s,t)+ 8 g^4 s u I^{(4)}_0 (s,u)+ 8 g^4 t u I^{(4)}_0 (t,u)
\end{align}
for the case of the $X_2^0 X_2^0 X_2^0 X_2^0$, ~$X_3^0 X_3^0 X_3^0 X_3^0$, or any of the $Y_q^0 Y_q^0 Y_q^0 Y_q^0$ amplitudes and 
\begin{align}
\mathcal{M}_{\utilde{\textbf{1}}}^{\textrm{Bose loops}} &= 8 g^4 (s^2+t^2 + 32 m^4) I^{(4)}_0 (s,t)+ 8 g^4 (s^2+u^2 + 32 m^4)I^{(4)}_0 (s,u) \nonumber \\
&+ 8 g^4 (t^2+u^2 + 32 m^4)I^{(4)}_0 (t,u) \\
\mathcal{M}_{\utilde{\textbf{1}}}^{\textrm{Fermi loops}} &= 8 g^4 (s t - 32 m^4) I^{(4)}_0 (s,t)+ 8 g^4 (s u - 32 m^4) I^{(4)}_0 (s,u) \nonumber \\ &+ 8 g^4 (t u -32 m^4) I^{(4)}_0 (t,u)
\end{align}
for the $X_1^0 X_1^0 X_1^0 X_1^0$ amplitude. In either case, summing the contributions and using the kinematic relation $s+t+u=0$ gives Eqn. 3.15 below. Although the individual components of the $X_1^0 X_1^0 X_1^0 X_1^0$ amplitude calculation had non-trivial mass dependence, this dependence cancels out in the final result, as we argued it must for all amplitudes in the massless sector of the theory. We find that Eqn. 4.9 of \cite{KM} should read
\begin{align}
\mathcal{M}^{4 \times X_2^0} &= 4 g^4 (s^2+t^2+u^2) \Big( I^{(4)}_0 (s,t)+I^{(4)}_0 (s,u)+I^{(4)}_0 (t,u) \Big).
\end{align}

Finally, we compute an example of a one-loop scattering amplitude with massive external states. For simplicity, we study the $X_2^0 X_2^0 X_2^+ X_2^-$ amplitude. It took somewhat more effort to check this amplitude since all previous work has focused exclusively on the massless sector of the theory. We performed all calculations both in unitary and 't Hooft-Feynman gauge. This check, however, is not sensitive to problems originating from the fermionic box graphs. It was useful to also keep track of all triangle diagrams and explicitly show that no scalar triangle integrals appear in the final result. The FeynArts \cite{KEM,H} package (as an add-on for FeynCalc) was used to generate the Feynman diagrams. We find
\begin{align}
\mathcal{M}^{X_2^0 X_2^0 X_2^+ X_2^-} = 2 g^4 (s^2+(t+m^2)^2+(u+m^2)^2)\Big( I^{(3)}_{2~ \textrm{hard}}(s,t)+I^{(3)}_{2 ~\textrm{hard}}(s,u)\Big),
\end{align}
where, as before, the scalar box integrals $I^{(3)}_{2 ~\textrm{hard}}(s,t)$ and $I^{(3)}_{2 ~\textrm{hard}}(s,u)$ are defined in Appendix A.

 It is remarkable that, apart from a factor of two, the $X_2^0 X_2^0 X_2^0 X_2^0$ and the $X_2^0 X_2^0 X_2^+ X_2^-$ amplitudes both have exactly the same coefficient structure. This can be easily seen by comparing the kinematics of the two cases. For $X_2^0 X_2^0 X_2^0 X_2^0$ we have
\begin{align}
2 k_3 \cdot k_4 = s && 2 k_1 \cdot k_3 = u && 2 k_1 \cdot k_4 = t
\end{align}
and for $X_2^0 X_2^0 X_2^+ X_2^-$, provided we choose $k_1^2=-m^2$,\\$k_2^2=-m^2$,~$k_3^2=0$,~ and $k_4^2=0$, we have
\begin{align}
2 k_3 \cdot k_4 = s && 2 k_1 \cdot k_3 = u+m^2 && 2 k_1 \cdot k_4 = t+m^2.
\end{align}

Upon examining the form of Eqns. 3.15 and 3.16, we see that the scattering amplitudes for these two processes are very similar. The fact that Eqns. 3.15 and 3.16 are so similar suggests that the technical complications arising from the inclusion of massive external states in $\mathcal{N}=4$ scattering amplitudes are not a serious issue and that a systematic analysis of our model's S-matrix should be straightforward to carry out.
\section{Summary and Future Directions}
~~~~~~~In this paper, we calculated a number of one-loop four-point scattering amplitudes in the Coulomb phase of $\mathcal{N}=4$ super Yang-Mills theory, both in the massless and massive sectors of the theory. Specifically, we studied the four photon ($A_{\mu}^0 A_{\nu}^0 A_{\rho}^0 A_{\sigma}^0$), two photon two massless scalar ($A_{\mu}^0 A_{\nu}^0 X_2^0 X_2^0$), four massless scalar ($X_2^0 X_2^0 X_2^0 X_2^0$), and two massless scalar two massive scalar ($X_2^0 X_2^0 X_2^+ X_2^-$) amplitudes. The results are all interesting and extremely simple, but most striking is the similarity between the amplitudes for $X_2^0 X_2^0 X_2^0 X_2^0$ and $X_2^0 X_2^0 X_2^+ X_2^-$. The complete Lagrange density of the model is given in Appendix B. The expressions given there should be extremely useful if one wishes to perform more involved calculations (\textit{e.g.} higher point or higher loop) in the model. 

The primary motivation for this work was the surprising fact that $\mathcal{N}=4$ scattering amplitudes with massive external states appear to have been completely neglected, in stark contrast to their massless counterparts. The $SU(2) \rightarrow U(1)$ model presented here should be useful for the testing and development of supersymmetric Ward identities in the massive sector of the model, continuing in the spirit of \cite{ST}. 

It may turn out that the massless sector of this $SU(2) \rightarrow U(1)$ $\mathcal{N}=4$ model actually  has more to offer, despite already being fairly well-understood. In their recent paper \cite{AM}, Alday and Maldacena introduce an infrared regulator for their one-loop four gluon scattering amplitude at strong coupling which amounts to considering an $SU(N) \rightarrow U(1)^{N-1}$ generalization of our model. The idea is that, when one goes to a generic point in the moduli space, the propagator denominators will be shifted as $(l-k_i)^2 \rightarrow (l-k_i)^2 + m_r^2$, eliminating the need for any dimensional regularization in the loop integrals. They argue that the color ordering of the planar $SU(N)$ four gluon amplitude is preserved when the gauge group is broken to $U(1)^{N-1}$.

In fact, the four photon amplitude studied in this paper is just the weak-coupling analog of the planar four gluon amplitude Alday and Maldacena calculated at strong coupling. It is easy to see that, if we kept only planar box graphs in the calculation of our four photon amplitude \footnote{The planar graphs are those which contribute to the coefficient of $I_0^{(4)}(s,t)$}, we would find, apart from constants, the same expression derived for ordinary $\mathcal{N}=4$ in \cite{BDDK}, but in the $(l-k_i)^2 \rightarrow (l-k_i)^2 + m_r^2$ regularization scheme described above, rather than dimensional regularization. It would be an interesting exercise to iterate the photon scattering amplitude to higher loop order, and attempt to compare the weak and strong coupling results for $\mathcal{N}=4$ four gluon scattering in this alternative IR regularization scheme. It would also be useful to check that other known iterative weak-coupling relations (see \textit{e.g.} \cite{ABDK,BDS}) hold in this alternative scheme as well.
\newpage
\acknowledgments
RMS would like to thank P.M. Chesler and L.G. Yaffe for useful discussions at the outset of this work. RMS is very grateful to M.J. Strassler for his ongoing support and for his critical reading of the draft. RMS thanks A.P. Sorini for his kind help in producing the figures. RMS would also like to thank M.E. Peskin for gracious hospitality at SLAC during the final phase of this work. Finally, RMS extends a very special thanks to L.J. Dixon for many illuminating discussions, helpful suggestions, and comments on the draft, without which the completion of this work would not have been possible.
\newpage
\appendix
\section{Scalar Box Integrals}
~~~~~~~In this appendix we define the scalar box integrals which arise in our one-loop calculations. All the Feynman integrals carry a superscript which specifies the number of internal lines which have mass $m$ 
and a subscript which specifies the number of massive external legs\footnote{For the subscripts, we follow the standard nomenclature outlined in \cite{BDKrev}; 
the label ``2 hard'' means that the process under consideration has its two massive external states adjacent, as opposed to being seperated on either side by a massless external state.
That situation would be labeled ``2 easy.''}. We begin with the four internal mass, zero external mass box integrals, $I_0^{(4)}(s,t)$,~$I_0^{(4)}(s,u)$, and $I_0^{(4)}(t,u)$. In what follows, we only discuss $I_0^{(4)}(s,t)$. The other two master integrals are easily derived through crossing symmetry relations.
\begin{align}
I_0^{(4)}(s,t) & = \int \frac{d^4 l}{(2 \pi)^4} \frac{1}{\big(l^2+m^2\big) \big((l-k_1)^2+m^2\big) \big((l-k_1-k_2)^2+m^2\big) \big((l+k_4)^2+m^2\big)},
\end{align}
where $k_1^2=k_2^2=k_3^2=k_4^2=0$. The labels $(s,u)$, $(s,t)$, and $(t,u)$ are, obviously, somewhat arbitrary. What we mean here is that the above box integrals have the following standard Feynman parametrization:
\begin{align}
I_0^{(4)}(s,t) & = \frac{i}{16\pi^2} \int_0^1 d x_1 
\int_0^{1-x_1} d x_2 \int_0^{1-x_1-x_2} d x_3 \frac{1}{\big(s x_2 (1-x_1-x_2-x_3) + t x_1 x_3 + m^2\big)^2}.
\end{align}
The explicit forms of $I_0^{(4)}(s,t)$,~$I_0^{(4)}(s,u)$, and $I_0^{(4)}(t,u)$ can be deduced from the expression found in Appendix A of \cite{BM}.

We now turn to the integrals, $I_{2~\textrm{hard}}^{(3)}(s,t)$ and $I_{2~\textrm{hard}}^{(3)}(s,u)$, which show up in the calculation of the $X_2^0 X_2^0 X_2^+ X_2^-$ scattering amplitude. Again, we only deal with $I_{2~\textrm{hard}}^{(3)}(s,t)$, as $I_{2~\textrm{hard}}^{(3)}(s,u)$ is easily obtained by crossing. In this case, we employ dimensional regularization, since only three of the internal lines are massive and we therefore expect IR divergences. We have 
\begin{align}
I_{2~\textrm{hard}}^{(3)}(s,t) & = \int \frac{d^{4-2 \epsilon} l}{(2 \pi)^{4-2 \epsilon}} \frac{1}{\big(l^2+m^2\big) \big(l-k_1\big)^2 \big((l-k_1-k_2)^2+m^2\big) \big((l+k_4)^2+m^2\big)}
\end{align}
where, as in Section Three, $k_1^2= -m^2$,~$k_2^2= - m^2$, and $k_3^2=k_4^2=0$. The explicit form of these master integrals follow from the result given in Appendix A of \cite{BKNS}.
\section{The Lagrange Density}
In this appendix we present the complete form of our model's Lagrange density. We do not derive the Feynman rules, since FeynArts does this when given the Lagrange density as an input and $\mathcal{L}$ is in any case a more basic object. We begin with the quadratic terms first discussed in Section Two,
\begin{align}
\mathcal{L}_{FP;2}=& -\frac{1}{2} A_{\mu}^0 \Big(-g^{\mu \nu} \partial^2+\partial^{\mu}\partial^{\nu} \Big( 1-\frac{1}{\xi}\Big) \Big)A_{\nu}^0-A_{\mu}^+ \Big( -g^{\mu \nu} \partial^2+\partial^{\mu}\partial^{\nu} \Big(1-\frac{1}{\xi}\Big)+m^2 g^{\mu \nu}\Big)A_{\nu}^- \nonumber \\& -X_1^+\Big(-\partial^2+ \xi m^2\Big)X_1^- -\frac{1}{2} X_1^0\Big(-\partial^2\Big)X_1^0  -\bar{c}\Big(-\partial^2+ \xi m^2\Big)c-\frac{1}{2} \bar{c}^0\Big(-\partial^2\Big)c^0
\end{align}
and all other pieces of the Lagrange density needed to determine the propagators of the fields:
\begin{align}
\mathcal{L}_{\bar{\psi}\psi} &= -\bar{\psi}^0_i \slashed{\partial} \psi^0_i - \utilde{\bar{\Phi}}_i \slashed{\partial} \utilde{\Phi}_i-m \big( \utilde{\bar{\Phi}}_1 \utilde{\Phi}_1 - \utilde{\bar{\Phi}}_2 \utilde{\Phi}_2 + \utilde{\bar{\Phi}}_3 \utilde{\Phi}_3  - \utilde{\bar{\Phi}}_4 \utilde{\Phi}_4 \big) \\
\mathcal{L}_{\chi \chi} &= - \frac{1}{2} X_2^0 \Big( -\partial^2\Big) X_2^0 -  X_2^+ \Big( -\partial^2+m^2 \Big) X_2^- - \frac{1}{2} X_3^0 \Big( -\partial^2\Big) X_3^0 \nonumber \\& -  X_3^+ \Big( -\partial^2+m^2 \Big) X_3^- -\frac{1}{2} Y_q^0 \Big( -\partial^2\Big) Y_q^0 -Y_q^+ \Big( -\partial^2+m^2 \Big) Y_q^- .
\end{align}

Next, we list all the cubic interaction terms in the purely bosonic sector present after spontaneous symmetry breaking and Faddeev-Popov gauge-fixing: 
\begin{align}
\mathcal{L}_{FP;3} &= -\xi m g X_1^0 \bar{c} c+\xi m g X_1^- \bar{c} c^0 -i g \bar{c} \partial^{\mu} \big(A^0_{\mu} c\big)+i g \bar{c} \partial^{\mu} \big(A_{\mu}^- c^0\big)+i g \bar{c}^0 \partial^{\mu} \big(A^+_{\mu} c\big) + \textrm{h.c.} \\
\mathcal{L}_{\chi A A} &= m g X_1^+ A_{\mu}^0 A^{-~\mu}+m g X_1^- A_{\mu}^0 A^{+~\mu}-2 m g X_1^0 A_{\mu}^+ A^{-~ \mu} \\
\mathcal{L}_{\chi 
\chi A} &=  i g A^-_{\mu} \big(X_p^+ \partial^{\mu} X_p^0-X_p^0 \partial^{\mu} X_p^+ \big) + i g A^-_{\mu} \big(Y_q^+ \partial^{\mu} Y_q^0-Y_q^0 \partial^{\mu} Y_q^+\big) \nonumber \\
&+ i g A^+_{\mu}\big(X_p^0 \partial^{\mu} X_p^--X_p^- \partial^{\mu} X_p^0\big) + i g A^+_{\mu}\big(Y_q^0 \partial^{\mu} Y_q^--Y_q^- \partial^{\mu} Y_q^0\big) \nonumber \\
&+ i g A^0_{\mu}\big(X_p^- \partial^{\mu} X_p^+-X_p^+ \partial^{\mu} X_p^-\big) + i g A^0_{\mu}\big(Y_q^- \partial^{\mu} Y_q^+-Y_q^+ \partial^{\mu} Y_q^-\big) \\
\mathcal{L}_{A A A} &= i g \partial_{\mu} A_{\nu}^+ \big(A^{0 ~\mu} A^{- ~\nu}- A^{0~ \nu} A^{- ~\mu}\big)+i g \partial_{\mu} A_{\nu}^0 \big(A^{+ ~\nu} A^{- ~\mu}- A^{+~ \mu} A^{- ~\nu}\big) \nonumber \\
& +i g \partial_{\mu} A_{\nu}^- \big(A^{0 ~\nu} A^{+ ~\mu}- A^{0~ \mu} A^{+ ~\nu}\big) \\
\mathcal{L}_{\chi 
\chi \chi} &= -m g X_1^0 X_{p'}^+ X_{p'}^- - m g X_1^0 Y_q^+ Y_q^-+ m g X_1^+ X_{p'}^0 X_{p'}^- + m g X_1^+ Y_q^0 Y_q^-+\textrm{h.c.}
\end{align}
Note that the sum over $p'$ in Eqn. B.8 excludes 1. The remaining three-point interactions involve the Fermionic fields:
\begin{align}
 \mathcal{L}_{A \bar{\psi} \psi} &= \frac{i g}{2}\utilde{\bar{\Phi}}_i \gamma^{\mu} \utilde{\Phi}_i A_{\mu}^0 - \frac{g}{\sqrt{2}} A_{\mu}^- \big(\utilde{\bar{\Phi}}_1 \gamma^{\mu} \psi^0_1+i \utilde{\bar{\Phi}}_1 \gamma^{\mu} \psi^0_2-\utilde{\bar{\Phi}}_2 \gamma^{\mu} \psi^0_1+i \utilde{\bar{\Phi}}_2 \gamma^{\mu} \psi^0_2\big) \nonumber \\ 
&- \frac{g}{\sqrt{2}} A_{\mu}^- \big(\utilde{\bar{\Phi}}_3 \gamma^{\mu} \psi^0_3+i \utilde{\bar{\Phi}}_3 \gamma^{\mu} \psi^0_4-\utilde{\bar{\Phi}}_4 \gamma^{\mu} \psi^0_3+i \utilde{\bar{\Phi}}_4 \gamma^{\mu} \psi^0_4\big) + \textrm{h.c.}
\end{align}

\begin{align}
&\mathcal{L}_{\chi \bar{\psi} \psi} = -\frac{g}{2} X_1^0 \big( \utilde{\bar{\Phi}}_1 \utilde{\Phi}_1 - \utilde{\bar{\Phi}}_2 \utilde{\Phi}_2 + \utilde{\bar{\Phi}}_3 \utilde{\Phi}_3  - \utilde{\bar{\Phi}}_4 \utilde{\Phi}_4\big)- \frac{g}{\sqrt{2}} X_{1}^- \big(i \utilde{\bar{\Phi}}_1 \psi^0_1- \utilde{\bar{\Phi}}_1  \psi^0_2+ i \utilde{\bar{\Phi}}_2 \psi^0_1+ \utilde{\bar{\Phi}}_2 \psi^0_2\big) \nonumber \\
&- \frac{g}{\sqrt{2}} X_{1}^- \big(i \utilde{\bar{\Phi}}_3 \psi^0_3- \utilde{\bar{\Phi}}_3  \psi^0_4+ i \utilde{\bar{\Phi}}_4 \psi^0_3+ \utilde{\bar{\Phi}}_4 \psi^0_4\big) + g X_2^0 \big( \utilde{\bar{\Phi}}_1 \utilde{\Phi}_4 - \utilde{\bar{\Phi}}_2 \utilde{\Phi}_3\big) \nonumber \\ 
& - \frac{g}{\sqrt{2}} X_{2}^- \big(i \utilde{\bar{\Phi}}_1 \psi^0_3+ \utilde{\bar{\Phi}}_1  \psi^0_4+ i \utilde{\bar{\Phi}}_2 \psi^0_3- \utilde{\bar{\Phi}}_2 \psi^0_4\big) + \frac{g}{\sqrt{2}} X_{2}^- \big(i \utilde{\bar{\Phi}}_3 \psi^0_1+ \utilde{\bar{\Phi}}_3  \psi^0_2+ i \utilde{\bar{\Phi}}_4 \psi^0_1- \utilde{\bar{\Phi}}_4 \psi^0_2\big) \nonumber \\
& + i g X_3^0 \big( \utilde{\bar{\Phi}}_1 \utilde{\Phi}_4 + \utilde{\bar{\Phi}}_2 \utilde{\Phi}_3\big)+ \frac{g}{\sqrt{2}} X_{3}^- \big( \utilde{\bar{\Phi}}_1 \psi^0_3- i \utilde{\bar{\Phi}}_1  \psi^0_4 - \utilde{\bar{\Phi}}_2 \psi^0_3- i \utilde{\bar{\Phi}}_2 \psi^0_4\big) \nonumber \\
&- \frac{g}{\sqrt{2}} X_{3}^- \big( \utilde{\bar{\Phi}}_3 \psi^0_1- i \utilde{\bar{\Phi}}_3  \psi^0_2 - \utilde{\bar{\Phi}}_4 \psi^0_1- i \utilde{\bar{\Phi}}_4 \psi^0_2\big) + \frac{i g}{2} Y_1^0 \big( \utilde{\bar{\Phi}}_1 \gamma^5 \utilde{\Phi}_1 - \utilde{\bar{\Phi}}_2 \gamma^5 \utilde{\Phi}_2- \utilde{\bar{\Phi}}_3 \gamma^5 \utilde{\Phi}_3+ \utilde{\bar{\Phi}}_4 \gamma^5 \utilde{\Phi}_4\big) \nonumber \\
&- \frac{g}{\sqrt{2}} Y_{1}^- \big(\utilde{\bar{\Phi}}_1 \gamma^5 \psi^0_1+ i \utilde{\bar{\Phi}}_1  \gamma^5 \psi^0_2+  \utilde{\bar{\Phi}}_2 \gamma^5 \psi^0_1-i  \utilde{\bar{\Phi}}_2 \gamma^5 \psi^0_2\big)
\nonumber \\
& + \frac{g}{\sqrt{2}} Y_{1}^- \big( \utilde{\bar{\Phi}}_3 \gamma^5 \psi^0_3+ i \utilde{\bar{\Phi}}_3  \gamma^5 \psi^0_4+  \utilde{\bar{\Phi}}_4 \gamma^5 \psi^0_3- i \utilde{\bar{\Phi}}_4 \gamma^5 \psi^0_4\big)  +i g Y_2^0 \big( \utilde{\bar{\Phi}}_1 \gamma^5 \utilde{\Phi}_3 - \utilde{\bar{\Phi}}_2 \gamma^5 \utilde{\Phi}_4\big) \nonumber \\
&- \frac{g}{\sqrt{2}} Y_{2}^- \big(\utilde{\bar{\Phi}}_1 \gamma^5 \psi^0_3+ i \utilde{\bar{\Phi}}_1  \gamma^5 \psi^0_4+  \utilde{\bar{\Phi}}_2 \gamma^5 \psi^0_3- i  \utilde{\bar{\Phi}}_2  \gamma^5 \psi^0_4\big) \nonumber \\
&- \frac{g}{\sqrt{2}} Y_{2}^- \big( \utilde{\bar{\Phi}}_3 \gamma^5 \psi^0_1+ i \utilde{\bar{\Phi}}_3 \gamma^5 \psi^0_2+ \utilde{\bar{\Phi}}_4 \gamma^5 \psi^0_1-i \utilde{\bar{\Phi}}_4 \gamma^5 \psi^0_2\big) + g Y_3^0 \big( \utilde{\bar{\Phi}}_1 \gamma^5 \utilde{\Phi}_3 + \utilde{\bar{\Phi}}_2 \gamma^5 \utilde{\Phi}_4\big) \nonumber\\
&+ \frac{g}{\sqrt{2}} Y_{3}^- \big(i \utilde{\bar{\Phi}}_1 \gamma^5 \psi^0_3- \utilde{\bar{\Phi}}_1  \gamma^5 \psi^0_4-i \utilde{\bar{\Phi}}_2 \gamma^5 \psi^0_3- \utilde{\bar{\Phi}}_2 \gamma^5 \psi^0_4\big)
\nonumber \\
&- \frac{g}{\sqrt{2}} Y_{3}^- \big(i \utilde{\bar{\Phi}}_3 \gamma^5 \psi^0_1- \utilde{\bar{\Phi}}_3 \gamma^5 \psi^0_2- i \utilde{\bar{\Phi}}_4 \gamma^5 \psi^0_1- \utilde{\bar{\Phi}}_4 \gamma^5 \psi^0_2\big) 
+ \textrm{h.c.}
\end{align}

Finally, we list the four-point interactions:

\begin{align}
\mathcal{L}_{A A A A} = &g^2A_{\mu}^0 A_{\nu}^0 A^{-~\nu} A^{+~\mu} - g^2A_{\mu}^0 A^{0~\mu} A_{\nu}^- A^{+~\nu}+\frac{g^2}{2}A_{\mu}^- A^{-~\mu} A_{\mu}^+ A^{+~\mu}-\frac{g^2}{2}A_{\mu}^- A_{\nu}^- A^{+~\mu} A^{+~\nu}
\end{align}

\begin{align}
\mathcal{L}_{\chi \chi A A} = &\frac{g^2}{2} X_p^{+} X_p^{+} A_{\mu}^- A^{-~\mu} +\frac{g^2}{2} X_p^{-}X_p^{-}A_{\mu}^+ A^{+~\mu}- g^2 X_p^- X_p^+ A_{\mu}^0 A^{0~\mu}+ g^2 X_p^+ X_p^0 A_{\mu}^0 A^{-~\mu} \nonumber \\
&+ g^2 X_p^- X_p^0 A_{\mu}^0 A^{+~\mu}-g^2 X_p^0 X_p^0 A_{\mu}^- A^{+~\mu}-g^2 X_p^+ X_p^- A_{\mu}^- A^{+~\mu} \nonumber \\
&\frac{g^2}{2} Y_q^{+} Y_q^{+} A_{\mu}^- A^{-~\mu} +\frac{g^2}{2} Y_q^{-}Y_q^{-}A_{\mu}^+ A^{+~\mu}- g^2 Y_q^- Y_q^+ A_{\mu}^0 A^{0~\mu}+ g^2 Y_q^+ Y_q^0 A_{\mu}^0 A^{-~\mu} \nonumber \\
&+ g^2 Y_q^- Y_q^0 A_{\mu}^0 A^{+~\mu}-g^2 Y_q^0 Y_q^0 A_{\mu}^- A^{+~\mu}-g^2 Y_q^+ Y_q^- A_{\mu}^- A^{+~\mu}
\end{align}
\begin{align}
\mathcal{L}_{\chi \chi \chi \chi} = &\sum_{l,k =1;~l < k}^3 \bigg( \frac{g^2}{2} X_l^{+~2} X_k^{-~2}+\frac{g^2}{2} X_l^{-~2} X_k^{+~2}-g^2 X_l^- X_l^+ X_k^{0~2}+g^2 X_l^+ X_l^0 X_k^0 X_k^- \nonumber\\
 &+g^2 X_l^- X_l^0 X_k^0 X_k^+-g^2 X_l^{0~2} X_k^+ X_k^- - g^2 X_l^- X_l^+ X_k^- X_k^+ + \frac{g^2}{2} Y_l^{+~2} Y_k^{-~2}+ \nonumber \\
& \frac{g^2}{2} Y_l^{-~2} Y_k^{+~2}-g^2 Y_l^- Y_l^+ Y_k^{0~2}+g^2 Y_l^+ Y_l^0 Y_k^0 Y_k^- +g^2 Y_l^- Y_l^0 Y_k^0 Y_k^+ -g^2 Y_l^{0~2} Y_k^+ Y_k^-\nonumber \\
& - g^2 Y_l^- Y_l^+ Y_k^- Y_k^+ \bigg)+\sum_{l,k = 1}^3\bigg( \frac{g^2}{2} X_l^{+~2} Y_k^{-~2}+\frac{g^2}{2} X_l^{-~2} Y_k^{+~2} -g^2 X_l^- X_l^+ Y_k^{0~2}\nonumber \\  
&  +g^2 X_l^+ X_l^0 Y_k^0 Y_k^- +g^2 X_l^- X_l^0 Y_k^0 Y_k^+-g^2 X_l^{0~2} Y_k^+ Y_k^- - g^2 X_l^- X_l^+ Y_k^- Y_k^+\bigg)
\end{align}
\section{Diagrammatic Representations of the Scattering Amplitudes}
~~~~~~~In this appendix, we give the Feynman diagrams for each one-loop scattering amplitude computed in Section Three. The program JaxoDraw \cite{BT} was used to draw the figures. Throughout, we give only the graphs that contribute to $(s,t)$ scalar box integrals, since those which contribute to the other channels are very similar. In what follows $p'$ runs over \{$2,3$\}, $q$ runs over \{$1,2,3$\}, and $i$ runs over \{$1,2,3,4$\}. In the massless sector, we omit charge conjugate graphs (graphs with all internal arrows reversed) for brevity. In calculating the amplitudes, one must simply remember to multiply the given diagrams by two. We also neglect Goldstone loops, since it is easy to recover the diagrammatic expansion in a general $R_\xi$ gauge once it is known in unitary gauge. All one needs to do is consider all possible ways in which internal $A_{\mu}^{\pm}$ lines could be replaced by Goldstone fields, $X_1^{\pm}$. 

\setcounter{bottomnumber}{2}
\setcounter{topnumber}{2}
\renewcommand{\bottomfraction}{.99}
\renewcommand{\textfraction}{0.01}
\newpage
$$\langle A_{\mu}^0 A_{\nu}^0 A_{\rho}^0 A_{\sigma}^0 \rangle$$
\begin{figure}[hbt]
\begin{narrow}{-.5 in}{0 in}
\resizebox{!}{45mm}{\includegraphics[width =\linewidth]{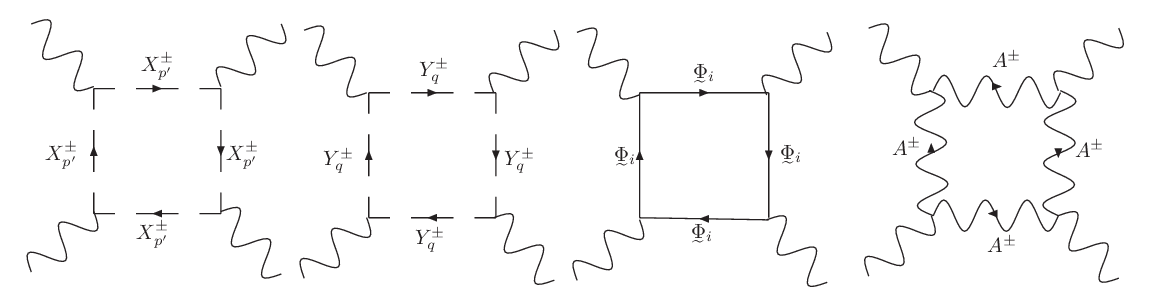}}
\end{narrow}
\end{figure}

$$\langle X_2^0 X_2^0 A_{\mu}^0 A_{\nu}^0 \rangle$$
\begin{figure}[hbt]
\begin{narrow}{-.7 in}{0 in}
\resizebox{!}{85mm}{\includegraphics[width =\linewidth]{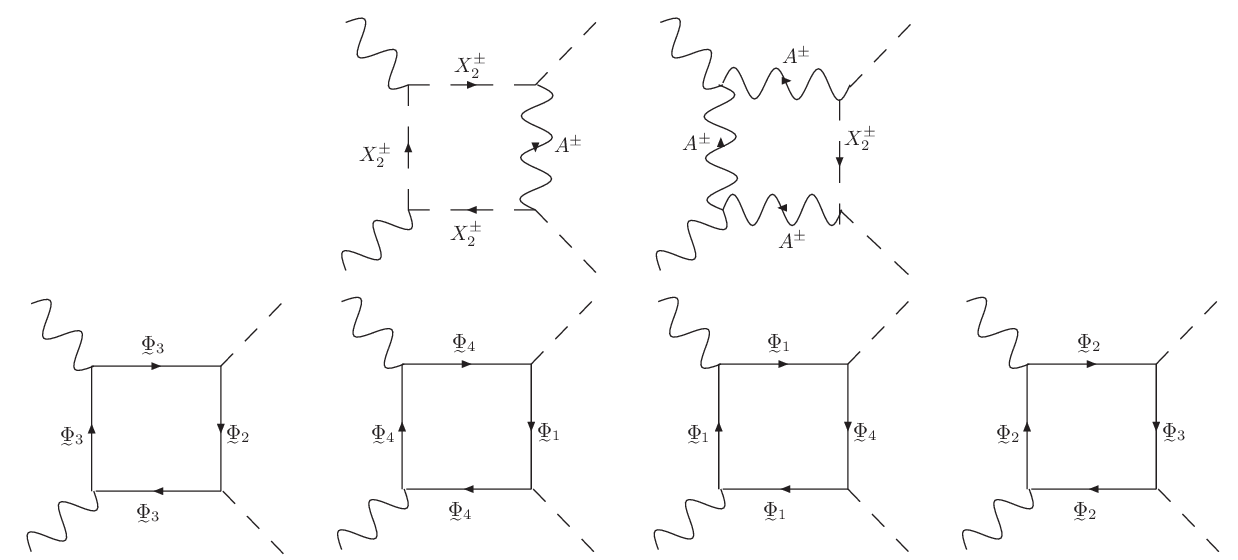}}
\end{narrow}
\end{figure}
\newpage
$$\langle X_1^0 X_1^0 X_1^0 X_1^0 \rangle$$
\begin{figure}[hbt]
\begin{narrow}{-.3 in}{0 in}
\resizebox{!}{40mm}{\includegraphics[width =\linewidth]{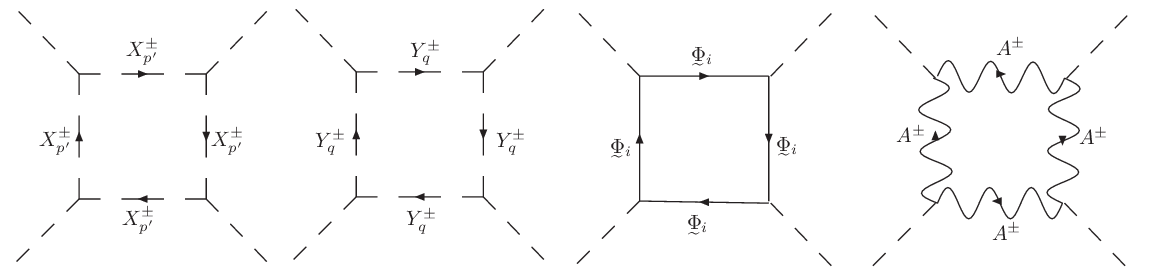}}
\end{narrow}
\end{figure}
$$\langle X_2^0 X_2^0 X_2^0 X_2^0 \rangle$$
\begin{figure}[hbt]
\begin{narrow}{-.5 in}{0 in}
\resizebox{!}{85mm}{\includegraphics[width =\linewidth]{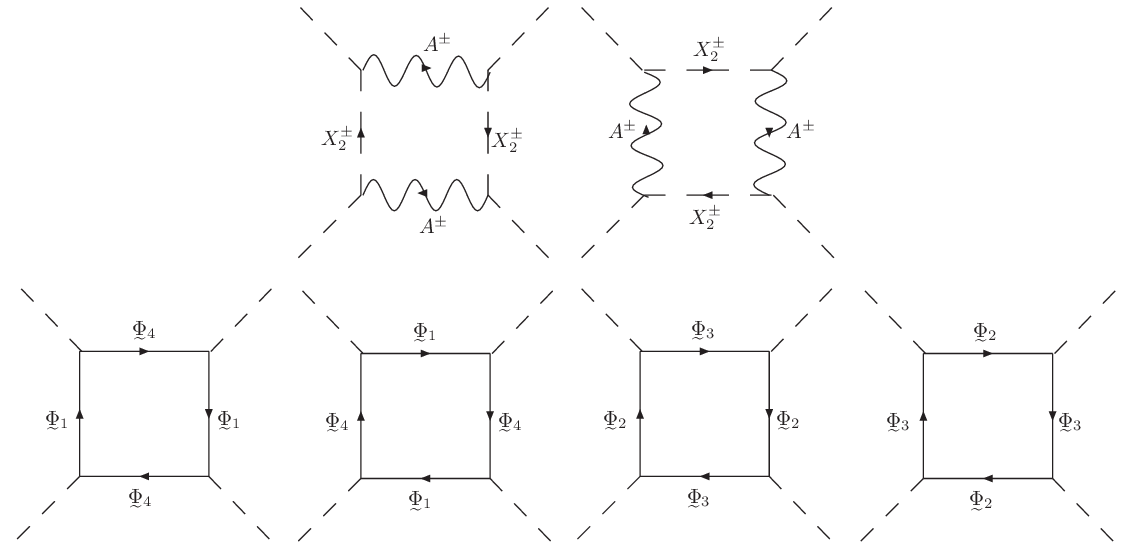}}
\end{narrow}
\end{figure}
\newpage
$$\langle X_2^0 X_2^0 X_2^+ X_2^- \rangle$$
\begin{figure}[hbt]
\begin{narrow}{-.5 in}{0 in}
\resizebox{!}{115mm}{\includegraphics[width =\linewidth]{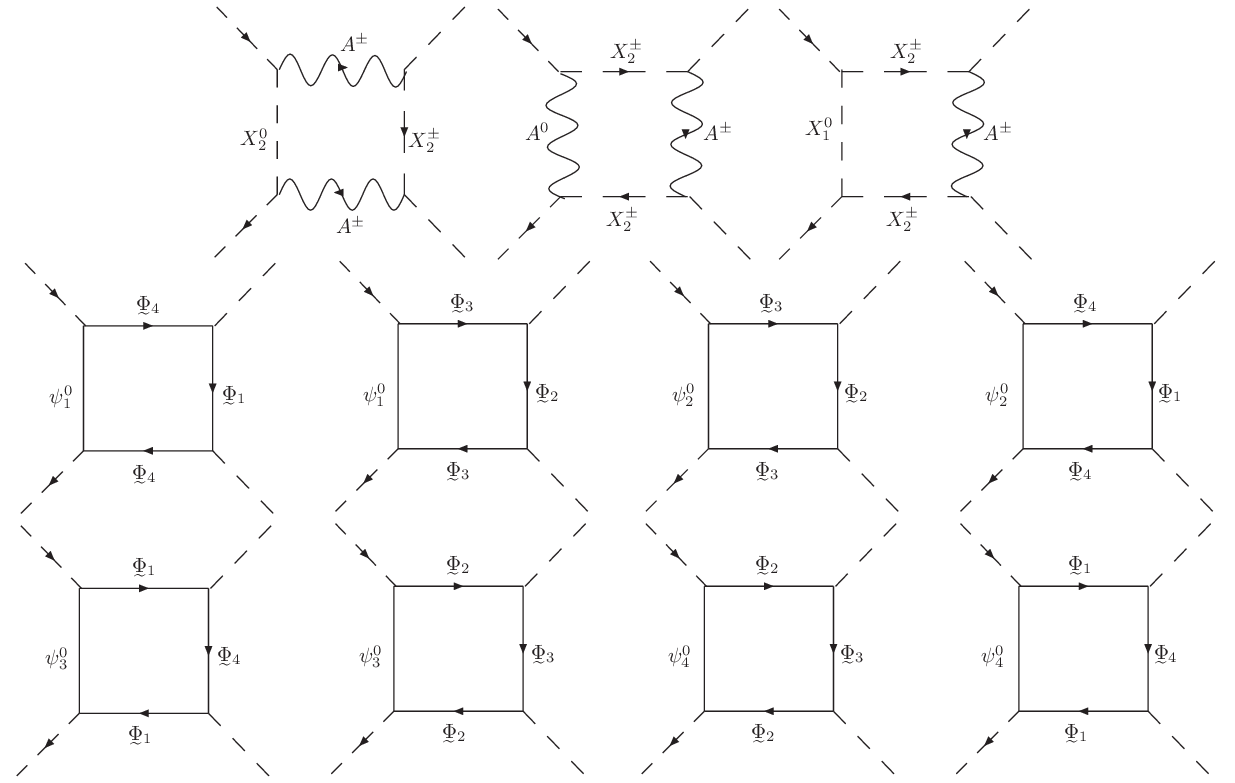}}
\end{narrow}
\end{figure}

\end{document}